\documentclass{aa}  

\usepackage{graphicx}
\usepackage{txfonts}
\usepackage{color}
\usepackage{wasysym}
\usepackage{natbib}

\begin{document}
    \title{The essential elements of dust evolution:}
    \subtitle{A viable solution to the interstellar oxygen depletion problem?}

    \author{A.P. Jones
           \and
           N. Ysard 
                    }

    \institute{Institut d'Astrophysique Spatiale, CNRS / Universit\'e Paris-Sud, Universit\'e Paris-Saclay, B\^atiment 121, Universit\'e Paris-Sud, 91405, Orsay Cedex, France\\
               \email{ Anthony.Jones@ias.u-psud.fr}
              }

    \date{Received ?, 2019; accepted ?, 2019}


   \abstract
{There remain many open questions relating to the depletion of elements into dust, {\it e.g.}, exactly how are C and O incorporated into dust in dense clouds and, in particular, what drives the disappearance of oxygen in the denser interstellar medium? }
{This work is, in part, an attempt to explain the apparently anomalous incorporation of O atoms into dust in dense clouds. }
{We re-visit the question of the depletion of the elements incorporated into the carbonaceous component of interstellar dust, {\it i.e.}, C, H, O, N and S, in the light of recent analyses of the organics in comets, meteorites and interplanetary dust particles. }
{We find that oxygen could be combined with $\approx 10$ to $20$\,\% of the carbon in the dust in dense regions in the form of a difficult to observe, organic carbonate, ($^{\rm -O}_{\rm -O}\hspace{-0.1cm}>${\tiny{C$=$O}}), which could explain the unaccounted for $170-255$\,ppm oxygen depletion.}
{We conclude that, while C, O and N atoms are depleted into an amorphous a-C:H:O:N phase, we posit that a significant fraction of C and O atoms could be sequestered into an organic carbonate, which provides a viable solution to the oxygen depletion problem. Further, the thermal or photolytic decomposition of this carbonate may have a bearing on the formation of CO$_2$ in the ISM. }


    \maketitle
%

\section{Introduction}

Observations clearly show that dust in the diffuse interstellar medium (ISM) is not the same everywhere \citep[{\it e.g.},][]{2007ApJ...663..320F,2009ApJ...699.1209F,2011A&A...536A..24P,2011A&A...536A..25P,2014ApJ...780...10L,2014ApJ...783...17L,2015A&A...577A.110Y,2015ApJ...811..118R,2017ApJ...834...63R,2017ApJ...851..119R,2017ApJ...846...38L,2018A&A...616A..71R,2018ApJ...862...49N,2018ApJ...862..131M}, despite the widely-held view to the contrary. This crucial fact has driven a detailed consideration of how and where dust evolves in the diffuse and dense ISM \citep[{\it e.g.},][]{2003A&A...398..551S,2008A&A...479..453Z,2008A&A...490..665P,2008AJ....136..919B,2013A&A...558A..62J,2013ApJ...773...30W,2014A&A...565L...9K,2014P&SS..100...32W,2015A&A...579A..15K,2015ApJ...811...38W, 2016A&A...588A..44Y,2016RSOS....360224J,2016ARep...60..669M}
and how it evolves in response to the local physical conditions, {\it e.g.}, gas temperature and density, interstellar radiation field (ISRF). This is particularly important in photo-dissociation regions \citep[PDRs, {\it e.g.},][]{2008A&A...491..797C,2010A&A...518L..96A,2012A&A...541A..19A,2012A&A...542A..69P,2015A&A...577A..16P,2018A&A...615A.129J} and proto-planetary discs \citep[{\it e.g.},][]{2017A&A...599A..80K,2018A&A...618L...1K,2019A&A...569A.100B}.  

Here we reconsider some crucial aspects of dust evolution in the transitions between low and high density quiescent regions ({\it i.e.}, the diffuse ISM and dense clouds) and in the high-density excited regions associated with star and planet formation. The exact nature of dust in a given environment still remains very much an open question and, in particular, how its evolution in a given environment depends upon gas-grain-ISRF interactions. These dust evolution processes operate primarily in both senses of the transitions between diffuse and dense clouds; through accretion and coagulation in the `forward' sense ({\it i.e.}, diffuse $\rightarrow$ dense) and photo-processing and disaggregation in star forming regions and PDRs in the `reverse' sense ({\it i.e.}, dense $\rightarrow$ diffuse), where the photolysis of dense cloud dust must seemingly re-populate the interstellar nanoparticle population in its evolution towards a diffuse ISM dust composition and structure. In the latter case the important processes that operate are the loss of ice mantles, most likely via photo-desorption \citep{1995P&SS...43.1311W}, and the de-volatilisation and fragmentation of aliphatic-rich a-C:H mantles \citep[{\it e.g.},][]{2014A&A...569A.119A,2015A&A...584A.123A} with embedded a-C nanoparticles \citep{2012A&A...542A..98J,2013A&A...558A..62J,2017A&A...602A..46J}.\footnote{Here and elsewhere in our work we use the term a-C(:H) to cover the whole family of aliphatic-rich and aromatic-rich materials, a-C:H and a-C, respectively.} 

In this work we are interested in the dominant solid phases that lock-up the principal dust constituent elements C, O, Si, Mg, Fe and S. Along the way we pay particular attention to the accretion and depletion of carbon and also focus on oxygen and the enigmatic problem of its disappearance into some as-yet unidentified material \citep{2009ApJ...700.1299J,2010ApJ...710.1009W,2016RSOS....360224J}. A solution to this problem, in the form of large, diffuse ISM ice grains with radii $\simeq 4\,\mu$m and O/H = 160\,ppm, was proposed by \cite{2015MNRAS.454..569W}, a proposition that requires a turnover timescale of $\approx 10^6 - 10^7$\,yr to replenish such grains from dense clouds. 

Perhaps related to the oxygen depletion problem is the observation that CO$_2$ is, apparently, rather abundant in the interstellar ices observed along lines of sight towards both quiescent and active star-forming clouds \citep[{\it e.g.},][]{1999ApJ...522..357G}. It has been shown that, under ISM-like conditions in the laboratory, CO molecules can react with O atoms on grains surfaces to form CO$_2$ \citep[{\it e.g.},][]{2001ApJ...555L..61R,2013A&A...559A..49M}. Further, interstellar chemistry models with diffusive grain surface reactions between CO and O can only reproduce the observed CO$_2$ abundances if the O atoms can efficiently migrate over the grain surfaces \citep{2001MNRAS.324.1054R}. Modelling has also shown that surface reactions between OH and CO are also capable of explaining CO$_2$ production in dark clouds \citep{2011ApJ...735...15G}. Nevertheless, it is apparent that the abundance determinations for CO$_2$ in interstellar ices are not without difficulties \citep{2015ApJ...808L..40G}. As a tangential issue to the main aims of this paper we briefly reconsider the possible origin of CO$_2$ in the ISM. 

In this paper we present 
our underlying dust philosophy (Section \ref{sect_dust_phil}) and on the basis of this  
consider the elemental abundances (Section \ref{sect_elements}), 
dust evolution (Section \ref{sect_dust_evol}),
the oxygen depletion problem (Section \ref{sect_O_problem}),
discuss our results (Section \ref{sect_discuss}),  
suggest experimental and observational strategies for testing the hypothesis (Section \ref{sect_elucidation}) 
and 
conclude (Section \ref{sect_concl}).

\section{The underlying dust philosophy}
\label{sect_dust_phil}

The underlying framework for this analysis is the The Heterogeneous dust Evolution Model for Interstellar Solids \citep[THEMIS,][]{2013A&A...558A..62J,2014A&A...565L...9K,2015A&A...579A..15K,2015A&A...577A.110Y,2017A&A...602A..46J},\footnote{http://www.ias.u-psud.fr/themis/} which was  constructed in a manner consistent with most of the critical constraints imposed by the direct and indirect sampling of dust, {\it i.e.}, by the knowledge gleaned from the analyses of meteorites and interplanetary dust particles (IDPs), interstellar dust in the solar system, the dust formed around evolved stars and in the ISM. The latter constraints, addressed by the THEMIS model, directly take into account two major aspects of dust formation processes, namely:
\begin{enumerate}
\item The bulk of the large grain mass is in Mg-rich amorphous silicate grains, a-Sil (with Fe/FeS nano-inclusions), which have grain sizes and compositions consistent with the pre-solar silicate grains extracted from meteorites and comets, and with observations of dust in the ISM \citep[{\it e.g.},][]{1994Sci...265..925B,2000ApJ...537..749C,2005A&A...444..187C,2006A&A...448L...1D,2011ApJ...738...78X,2012A&A...539A..32C,2014Sci...345..786W}. 
The isotopic compositions of these pre-solar grains trace their origins to formation in the circumstellar shells around evolved stars \cite[{\it e.g.},][]{2003M&PSA..38.5185M,2004ApJ...613L.149M,2004LPI....35.1593M,2004LPI....35.1675N,2004M&PSA..39.5158N,2007ApJ...656.1223N}.
\item  These large a-Sil grains are assumed to have amorphous carbon, a-C, mantles, as also are the large hydrogenated amorphous carbon grains, a-C:H, {\it i.e.}, in a-Sil$_{\rm Fe,FeS}$/a-C and a-C:H/a-C core/mantle structures, which are morphologically and compositionally typical of ISM dust growth through the accretion of gas phase carbon and the coagulation of carbonaceous nanoparticles onto their surfaces.\footnote{The viability of core/mantle grain structures was recently called into question using the tensile strength for water ice for the mantles \citep{2018arXiv181208391H,2019arXiv190206438H}. This does not apply to a-C mantles, which have tensile strengths considerably larger than that of ice.}
\end{enumerate}
These critical elements play a key role in driving our understanding of the origin, formation and evolution of dust \citep[{\it e.g.},][]{2011A&A...530A..44J,2014A&A...570A..32B} and were crucial in the construction and consolidation of the THEMIS dust model \citep{2012A&A...540A...1J,2012A&A...540A...2J,2012A&A...542A..98J,2013A&A...558A..62J,2014A&A...565L...9K,Faraday_Disc_paper_2014,2015A&A...577A.110Y,2015A&A...579A..15K,2016A&A...588A..44Y,2016RSOS....360221J,2016RSOS....360223J,2016RSOS....360224J,2017A&A...602A..46J,2018arXiv180410628J}.

This modelling philosophy appears to be generically consistent with recent analyses of GEMS-containing IDPs and the interpretation of those data as being mixed phases of a-Sil, Fe/Ni metal and sulphides, with a two component (aliphatic and aromatic) organic carbon mix \citep{2018PNAS..115.6608I}. The model is also coherent with the organic nano-globules and carbon-coated amorphous silicates observed in pre-solar meteoritic grains \citep{2002IJAsB...1..179N,2004E&PSL.224..431G,2006Sci...314.1439N,2006LPI....37.1455G,2008LPI....39.2391M,2009LPI....40.2260D,2009_CLS_Report_DeGregorio,2009AGUFM.P14A..02D,2009LPI....40.1130D,2015LPI....46.1609D,2016RSOS....360224J}.

\section{The elements}
\label{sect_elements}

A significant fraction ($\simeq 20-100$\,\%) of the elements C, O, Si, Mg and Fe are assumed to be locked up in a solid dust phase in the ISM, {\it i.e.}, interstellar dust. Table~\ref{tab_THEMIS_abunds} shows the typical abundances and depletions of these elements based upon the THEMIS dust modelling framework \citep{2013A&A...558A..62J,2014A&A...565L...9K,2015A&A...577A.110Y,2015A&A...579A..15K,
2017A&A...602A..46J}. Given that there exist different determinations of the cosmic abundance of oxygen, the O abundances of 490 and 575 ppm used here are taken from \cite{2009ARA&A..47..481A} and \cite{2008ApJ...688L.103P}, respectively, where (490:575) in Table~\ref{tab_THEMIS_abunds} indicates the possible range of the cosmic oxygen abundance, a span of $\sim 16$\,\%. 

Even though nitrogen is not usually thought of as an element important for dust, interstellar depletion studies indicate that it does exhibit low-level depletion in the diffuse ISM \citep{2009ApJ...700.1299J}. Further, and as it is an important element in meteoritic, interplanetary dust particle (IDP) and cometary organics, it does therefore merit consideration in dust evolution studies. Based on this reasoning, Table~\ref{tab_THEMIS_abunds} also indicates the possible incorporation of N and O heteroatoms,  at the $\sim 10$\,\% level,  relative to carbon, within the THEMIS a-C(:H) grain component, {\it i.e.}, as a-C(:H)[:O:N], where the square brackets are used to indicate minor (heteroatom) elements within a-C(:H). 

In general, the measurement uncertainties in elemental abundance determinations are not of great concern here because we are interested in larger-scale effects ($\gg 100$\,ppm) that trace global evolution trends in the chemical composition of the more volatile component of interstellar dust. Hence, and in order to demonstrate this, we consider the two alternative O abundance measurements shown in Table~\ref{tab_THEMIS_abunds} \citep[{\it i.e.}, 490 and 575\,ppm,][respectively]{2009ARA&A..47..481A,2008ApJ...688L.103P}.  

\begin{table}
\caption{Illustrative elemental abundances and depletions for dust-forming elements in the diffuse ISM, based upon the THEMIS dust model \citep{2017A&A...602A..46J}. The possible range of O abundances are indicated \citep[490:575,][respectively]{2009ARA&A..47..481A,2008ApJ...688L.103P}. The C abundance is taken from \cite{2009ApJ...700.1299J} and \cite{2012ApJ...760...36P}. Otherwise the abundances are from \cite{2009ApJ...700.1299J} and \cite{2014pacs.book...15P}. As explained in the text, $\sim 10$\,\% of O and N heteroatoms, relative to carbon, has been included within a-C(:H) mantles.}             
\label{tab_THEMIS_abunds}      
\centering                          
\begin{tabular}{c c c c c}        
\hline\hline \\[-0.25 cm]                 
              &  Total           & In dust &  Dust & Residual \\    
Element &      [ppm]      & [ppm]         &    phase &  [ppm] \\[0.1cm]    
\hline \\[-0.25 cm]                        
    O   &    (490:575) &        110 &        a-Sil         &     (360:445) \\      
         &                   &        20  &    a-C(:H)[:O:N]  &                  \\[0.1cm]      
    N   &     79          &        20  &    a-C(:H)[:O:N]  &        59       \\[0.1cm]      
    C   &     390        &        206 &        a-C(:H)     &        184     \\      
    Si  &     40          &        32   &        a-Sil          &          8       \\      
    Mg &     42         &        45   &        a-Sil          &       $-3$       \\      
    Fe  &     35         &        19   &        Fe,FeS      &         16     \\      
    S   &    15           &        3     &        FeS           &         12     \\[0.1cm]      
\hline  \\[-0.25 cm]                                 
\end{tabular}
\end{table}

\section{The essentials of dust evolution}
\label{sect_dust_evol}

\begin{figure}
\centering
\includegraphics[width=9cm]{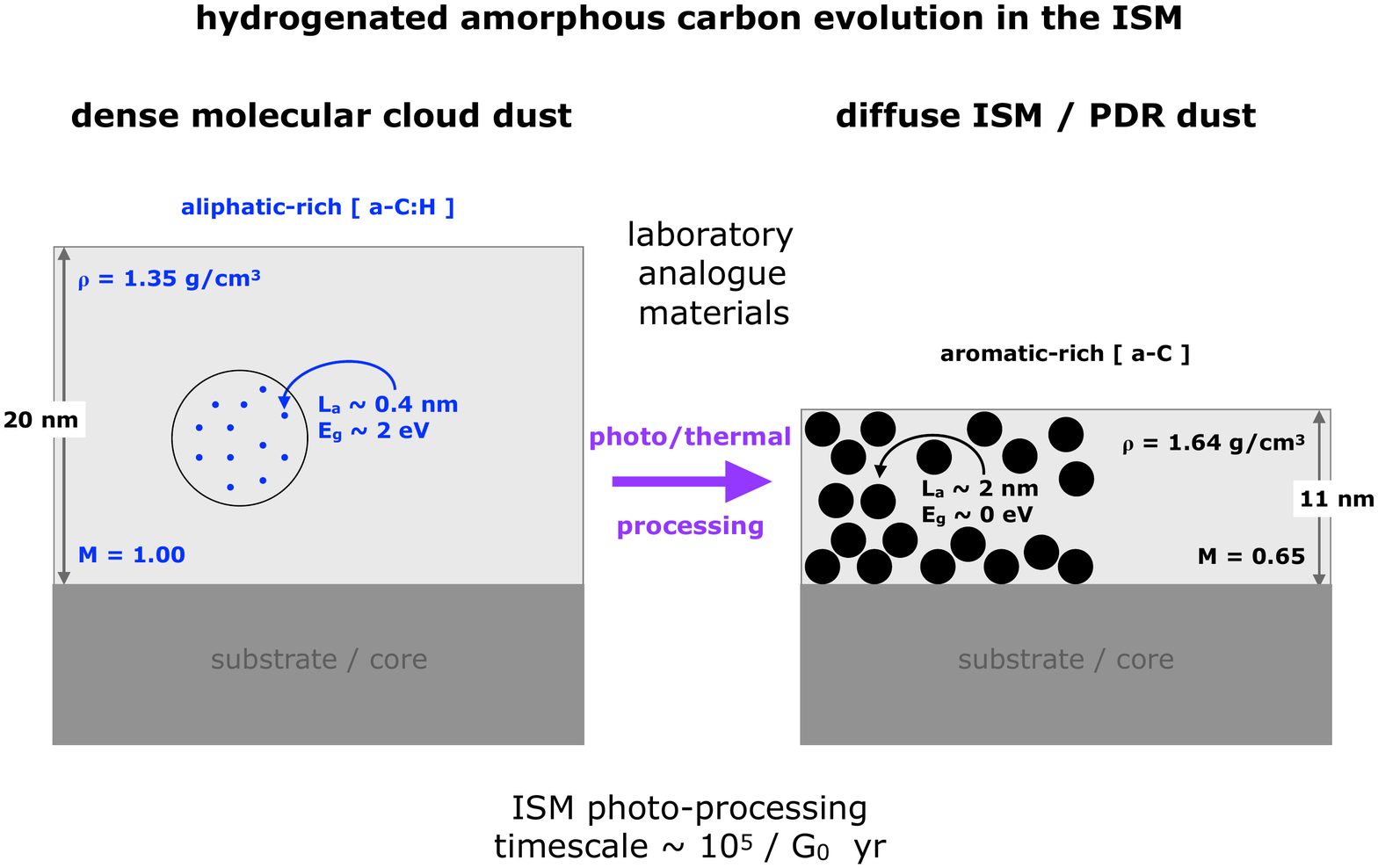}
\caption{A schematic summary of the main elements of laboratory experiments on a-C(:H) materials used as analogues for interstellar dust processing and evolution in the THEMIS dust model and based upon the experimental work of \cite{1984JAP....55..764S} and \cite{2011A&A...528A..56G}. Here {\sf M} represents the normalised mass of the deposited or mantle material.}
\label{fig_analogues}
\end{figure}

\begin{figure*}
\centering
\includegraphics[width=18cm]{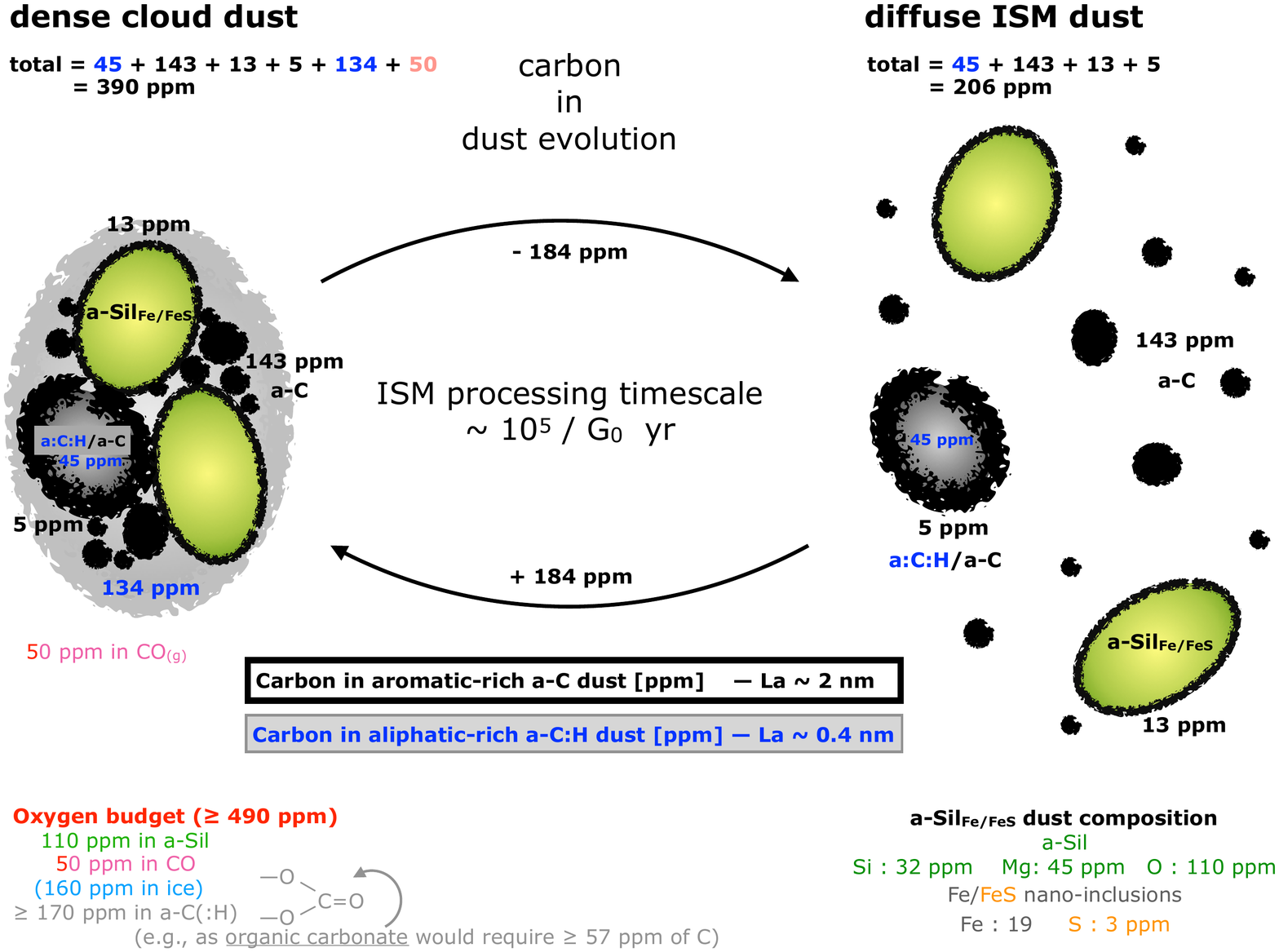}
\caption{A schematic view of the elemental abundance (re-)distribution for the THEMIS modelling approach to dust evolution in the transitions between dense clouds and the diffuse ISM. Ice mantles, which require $\simeq 160$\,ppm of oxygen \citep[{\it e.g.},][]{2010ApJ...710.1009W}, are not illustrated for clarity. The repartition of all the major dust-forming elements (O, C, Si, Mg, Fe and S) are indicated in ppm.}
\label{fig_general}
\end{figure*}

\begin{table}
\caption{Summary of the approximate physical conditions for the diffuse, translucent and dense cloud phases of the ISM using, in part, the relation $N_{\rm H} = 8.3 \times 10^{21}$\,cm$^{-2}\,E(B-V)$ \citep{2014ApJ...780...10L,2014ApJ...783...17L}.} 
\label{table}      
\centering                          
\begin{tabular}{l c c c l}        
\hline\hline \\[-0.25 cm]                 
 property & diffuse &  translucent &  dense   &  units \\[0.1 cm]    
\hline \\[-0.25 cm]                        
  $n_{\rm H}$    &   $10^0-10^2$  &  $10^2-10^5$     &  $> 10^5$    & cm$^{-3}$          \\[0.1 cm]      
  $T_{\rm gas}$ &   $\simeq 100$  &  $100-20$          &  $10-20$     & K                        \\[0.1 cm]      
  ISRF               &   $\simeq 1$      &   $< 1$               &   $\ll 1$         &  G$_0$            \\[0.1 cm]      
  $R_{\rm V}$    &  $\sim 3$          &   $\sim 3-5$        &  $\gtrsim 5$  &  ---                   \\[0.1 cm]      
  $A_{\rm V}$    &  $< 1$               &  $\sim 1-5$         &  $> 5$          &  mag.         \\[0.1 cm]      
   E(B-V)           &  $\lesssim 0.3$ &   $0.3 - 1$          &   $> 1$          &  mag.          \\[0.1 cm]      
   $N_{\rm H}$  &  $< 10^{21}$     &   $10^{21} - 10^{22}$          &  $> 10^{22}$   &  cm$^{-2}$   \\[0.1 cm]      
   f(H$_2$)        &   $\simeq 0$     &    $< 1$               &   $\simeq 1$   & ---                  \\[0.1 cm]      
 \hline   \\[-0.25 cm]                                 
\end{tabular}
\label{table_ISM}
\end{table}

Before considering dust evolution in the ISM it is perhaps useful to define the physical conditions of the different phases of the media within which this evolution takes place. To this end, and to aid the reader, Table~\ref{table_ISM} summarises the approximate physical conditions of the interstellar media under consideration here. 

Several outstanding issues in ISM studies would seem to be related to accretion effects, {\it e.g.}, the variable carbon abundance \citep{2012ApJ...760...36P}, the oxygen depletion problem \citep{2009ApJ...700.1299J,2010ApJ...710.1009W} and the unexplained disappearance of sulphur from the gas in dense regions. Of particular interest here is oxygen which depletes from the gas faster than can be accounted for by incorporation into a silicate/oxide dust or into an icy phase \citep{2009ApJ...700.1299J,2010ApJ...710.1009W}. 

The THEMIS dust modelling framework primarily considers dust evolution in terms of the accretion of C and H atoms, however, grain surfaces  in the ISM must also  accrete other abundant gas phase atoms, such as O, N, S, {\it etc.} Clearly, grain charge could play an important role during accretion, in that C and S are ionised  in the diffuse and translucent ISM and accretion onto positively-charged large grains could be Coulomb-hindered. However, the carbonaceous nanoparticles ($a \leqslant 3$\,nm), which provide $> 90$\,\% of the accreting grain surface in the THEMIS dust model \citep{2016RSOS....360221J} are likely to be predominantly neutral \citep[{\it e.g.},][]{2001ApJS..134..263W} and can therefore accrete positive ions, such as C$^+$ and S$^+$,  from the gas phase. Hence, there do not appear to be any obvious barriers to the accretion of any of the major dust forming elements in the diffuse ISM.

\subsection{Carbon accretion and evolution}
\label{sect_C_accretion}

In Fig.~\ref{fig_analogues} we present a schematic summary of some important laboratory results \citep{1984JAP....55..764S,2011A&A...528A..56G} that, within the framework of the THEMIS model, illustrate what are considered to be the extreme states of a-C(:H) evolution in the ISM \citep{2012A&A...542A..98J,2013A&A...558A..62J,Faraday_Disc_paper_2014,2016RSOS....360224J,2017A&A...602A..46J}. Namely, the H-rich and aliphatic-rich a-C:H formed by the accretion of C and H atoms from the gas within and on the borders of dense clouds where the ISRF is highly-attenuated. This material, under the effects of thermal- and/or UV photo-processing due to the local (and possibly enhanced) ISRF, will evolve towards an end-of-the-road H-poor and aromatic-rich a-C material \citep{2012A&A...542A..98J,2013A&A...558A..62J}. The essential elements of this a-C(:H) evolutionary scenario, starting with aliphatic-rich a-C:H, are:
\begin{enumerate}
\item A loss of the more volatile or polymeric matter, concomitant with an overall decrease of $\simeq 35$\,\% in mass \citep{1984JAP....55..764S}. 
\item A densification of the material from $\simeq 1.35$ to $\simeq 1.64$\,g\,cm$^{-3}$ \citep{1984JAP....55..764S}. 
\item An increase in the aromatic domain size, $L_a$, from $\simeq 0.4$\,nm to $\simeq 2$\,nm \citep{2011A&A...528A..56G} or to sizes that are limited by the particle dimensions \citep{2012A&A...542A..98J}. 
\item A decrease in the optical gap from $\simeq 2$\,eV to $\simeq 0$\,eV \citep{1984JAP....55..764S,2012A&A...540A...2J}.
\end{enumerate}
With this dust evolution scenario in mind it has been possible to estimate the a-C(:H) UV photo-processing ($E_{h\nu} \geqslant 10$\,eV) time-scale as of the order of $10^5/G_0$\,yr \citep{2012A&A...540A...2J,2012A&A...542A..98J,Faraday_Disc_paper_2014,2015A&A...581A..92J}, where $G_0$ is the scaling factor for the ISRF in the local Milky Way neighbourhood.

\begin{figure*}
\centering
\includegraphics[width=18cm]{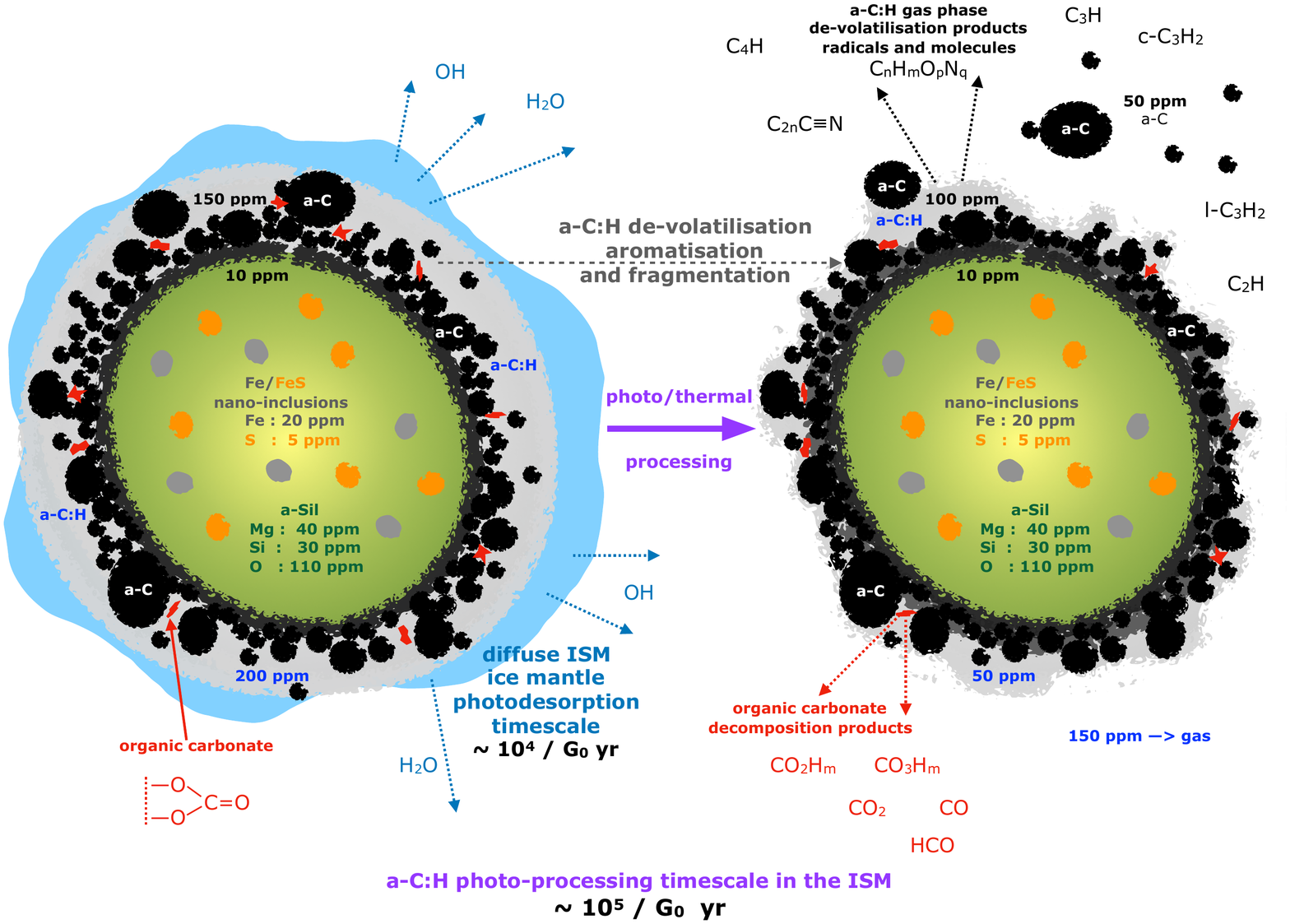}
\caption{A schematic view of the essentials of dust evolution in the intermediate transition states between lightly-processed dense cloud dust (left) and dust that is evolving towards that of the diffuse ISM dust (right). In this case the indicated elemental abundances are only illustrative order of magnitude indications.}
\label{fig_essentials}
\end{figure*}

Fig.~\ref{fig_general} gives a schematic view of the refractory, dust-forming elemental (re-)distribution for dust evolution, based upon the THEMIS model \citep{2013A&A...558A..62J,2017A&A...602A..46J,2016RSOS....360224J}. This figure summarises the depletion variations and the dust evolution between low-density and high-density interstellar media, including translucent clouds and the outer reaches of dense clouds where the onset of accretion and dust coagulation occurs. This schematic view does not include volatile ices in the diagram but they are accounted for, within the considered dust aggregates, in the oxygen budget in the lower left corner of the figure. Perhaps the most significant effect shown in Fig.~\ref{fig_general} is the exchange of a significant fraction of carbon ($\sim 50$\,\%) and oxygen ($\sim 70$\,\%) between the gas and dust phases in the transition between dense and diffuse clouds.\footnote{For now, but soon justified, we  assume that oxygen atoms are depleted into a-C:H:O mantles in a C:O stoichiometric ratio of 1:3, where the descriptor a-C:H:O indicates hydrogenated amorphous carbon material, a-C:H, with a significant oxygen atom content, {\it i.e.}, $> 10$\,\%.} In this evolutionary sense (dense to diffuse ISM) the mass loss from a-C:H mantles is $\sim 35$\,\%, , {\it i.e.}, $\equiv (184-50)/390$, if we only consider carbon atom loss from a-C(:H) and do not count the carbon in CO$_{\rm (g)}$ (50\,ppm), which is compatible with the 35\,\% mass loss seen in experiments \citep[Fig.\ref{fig_analogues} and ][]{1984JAP....55..764S}. For a-C:H:O mantle processing more than half of the mantle mass ($\sim 55$\,\%) is in O atoms, which would then dominate the $\sim 46$\,\%\footnote{{\it i.e.}, $((184-50) \times 12 + 170 \times 16)/ (340 \times 12 + 170 \times 16) = 0.46$.} mass loss from the a-C:H:O mantles, {\it i.e.}, considering only O and C atom loss, which still compares reasonably well with the 35\,\% mass loss of the experiments. Thus, it appears that the THEMIS-proposed dust evolutionary scenario (Fig.~\ref{fig_general}) is compatible with the observed interstellar C and O depletion variations and also with the measured evolution of a-C(:H) materials in the laboratory. 

It is likely that gas phase CO could also participate in this gas-dust mass exchange with CO$_{(g)}$$\downarrow$ being incorporated into dust mantles via accretion and also possibly by being chemically-sequestered into an organic (carbonate) phase by reaction with epoxide functional groups on grain surfaces in translucent clouds \citep[][see later]{2016RSOS....360224J}. 

The processes involved  in dust evolution operate over characteristic time-scales and in Fig.~\ref{fig_essentials} we show intermediate transition states, between dense cloud dust that is beginning to lose its ice mantles (left hand side of Fig.~\ref{fig_essentials}) and PDRs where the photo-processing of dense cloud dust is beginning to re-populate the interstellar nanoparticle population and is evolving towards a diffuse ISM dust composition and structure (right hand side of Fig.~\ref{fig_essentials}). The important processes shown in this figure are the loss of ice mantles, most likely via photo-desorption \citep{1995P&SS...43.1311W}, and the de-volatilisation \cite[{\it e.g.},][]{2014A&A...569A.119A,2015A&A...584A.123A} and fragmentation of aliphatic-rich a-C:H mantles with embedded a-C nanoparticles \citep{2016RSOS....360223J}.

\subsection{Oxygen accretion}
\label{sect_O_accretion}

We now consider in more detail the incorporation of O atoms into a-C:H:O materials (as mentioned above), as has been discussed from an observational \citep{2009ApJ...700.1299J,2010ApJ...710.1009W} and a theoretical point of view \citep{2010ApJ...710.1009W,2015MNRAS.454..569W,2016RSOS....360221J,2016RSOS....360223J,2016RSOS....360224J}.

\begin{figure}
\centering
\includegraphics[width=8cm]{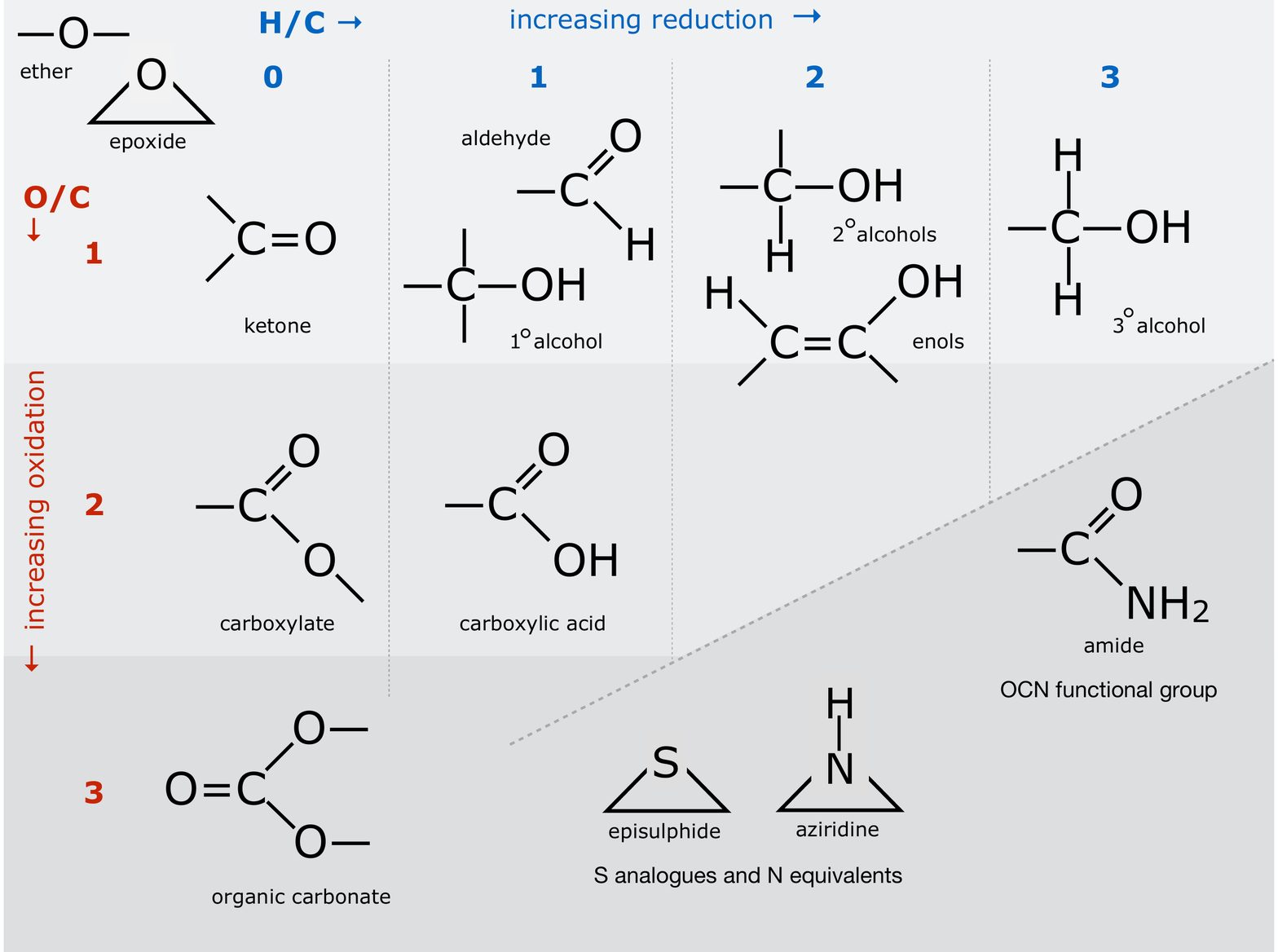}
\includegraphics[width=8cm]{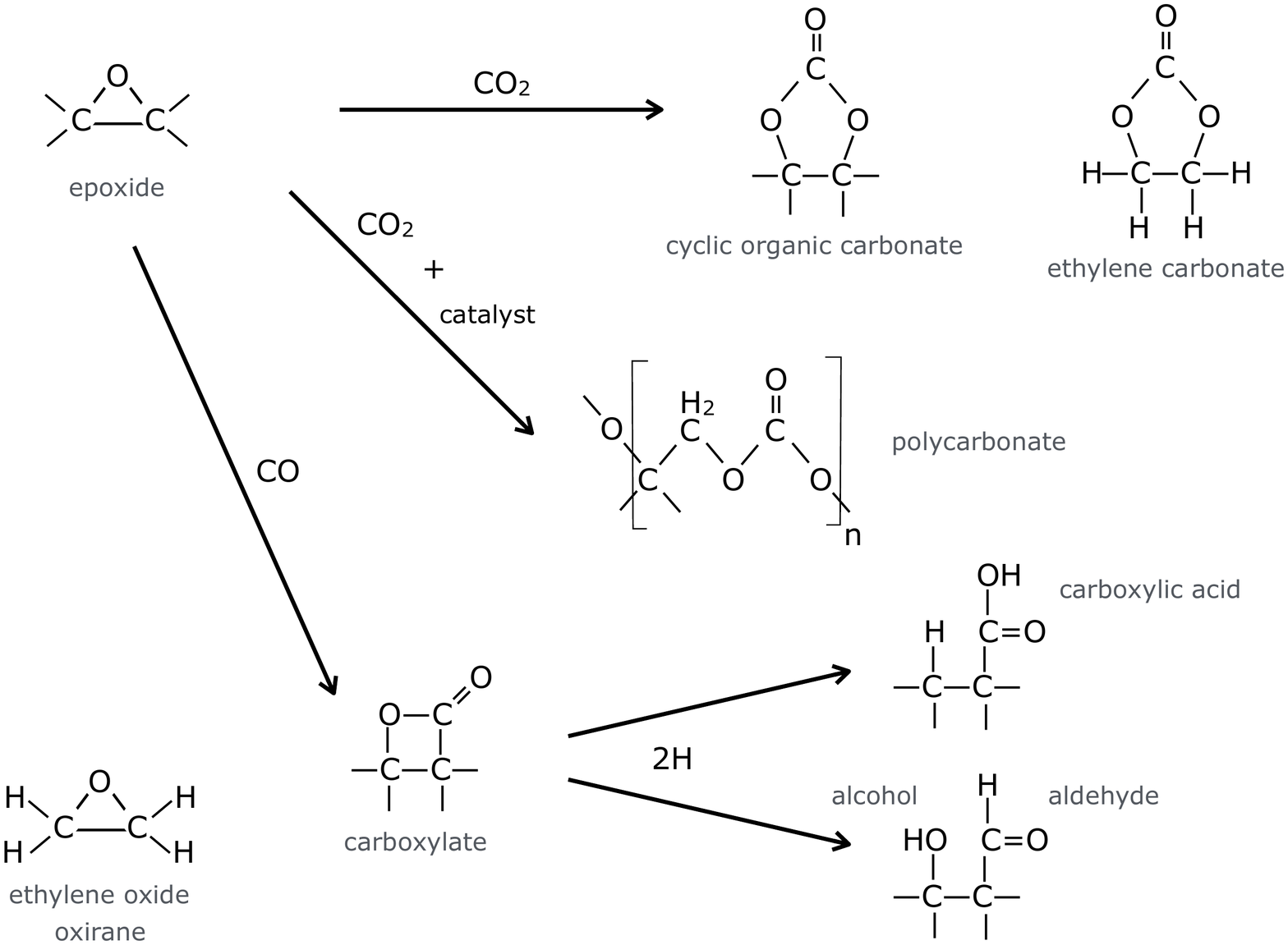}
\caption{The rich epoxide-driven carbonyl chemistry possible within a-C:H:O[:N:S] solids (upper panel): N atom reaction with epoxides and CO molecule reactions with aziridines could provide an efficient route to abundant amides \citep[][]{2012A&A...542A..98J}. The lower panel shows likely routes to a variety of organic carbonate and related structures.}
\label{fig_chemistry}
\end{figure}

\cite{2016RSOS....360224J} considered the formation of carbonyl functional groups (C=O) within a-C(:H) materials via O atom and CO molecule reactions with C$=$C bonds to form surface epoxide and carboxylate/carboxylic acid groups, respectively: the proposed chemistry is outlined in Fig.~\ref{fig_chemistry}, where the different organic functional groups mentioned in the following are illustrated (upper panel). Further, \cite{2016RSOS....360224J} proposed that gas phase CO$_2$ could react directly with surface epoxide, $>$C$_{-}^{\rm O}$C$<$, species and be sequestered into "organic carbonates" (lower panel in Fig.~\ref{fig_chemistry}), which would seem most likely to form cyclic species \citep{2016RSOS....360224J}. Thus it is possible that a significant fraction of the gas phase atoms and molecules (O, C, CO and CO$_2$) incident upon interstellar a-C(:H) grain surfaces in the tenuous or translucent ISM could end up in an oxygenated a-C:H material (a-C:H:O) rich in carbonyl bonds (C$=$O) in aldehyde, ketone, carbonate and amide functional groups, {\it e.g.}, $-$C$\leqslant^{\rm H}_{\rm O}$,  $>$C$=$O, $^{\rm -O}_{\rm -O}$$>$C$=$O and $-$C$\leqslant^{\rm NH_2}_{\rm O}$, respectively. Further, and in an analogous manner N and S atoms could also be incorporated during accretion to form aziridine, $>$C$_{-}^{\rm N}$C$<$, and episulphide, $>$C$_{-}^{\rm S}$C$<$, functional groups, which would react to form species containing imine and nitrile, {\it e.g.}, $>$C$=$N and $-$C$\equiv$N groups, and the sulphur analogues of aldehyde and ketone, {\it e.g.}, $-$C$\leqslant^{\rm H}_{\rm S}$ and  $>$C$=$S, respectively. It should in particular be noted that amide functional groups, incorporating both O and N heteroatoms, are particularly abundant in cometary volatiles, with formamide, H$-$C$\leqslant^{\rm NH_2}_{\rm O}$, being the most abundant molecule (at the 2\% level) after water \citep{2015Sci...349b0689G}. All of these findings and suppositions appear to be consistent with the analyses of GEMS-containing IDPs \citep[{\it e.g.},][]{2009LPI....40.1130D,2009LPI....40.2260D,2009AGUFM.P14A..02D,2018PNAS..115.6608I}.

\section{A viable solution to the O depletion problem}
\label{sect_O_problem}

The rapid removal of O atoms from the interstellar gas, in the transition from diffuse to translucent to dense clouds, faster than can be explained by its uptake into silicates, oxides, CO molecules or H$_2$O-dominated ice mantles remains a veritable  conundrum \citep[{\it e.g.},][]{2009ApJ...700.1299J,2010ApJ...710.1009W,2016RSOS....360224J}. The problem is that, whatever phase the oxygen depletes into, it must deplete into it to the level of $\simeq 170-255$\,ppm \citep[{\it e.g.},][]{2010ApJ...710.1009W}, corresponding to locking-up $\simeq 35-44$\,\% of the total available oxygen in this unknown material, which clearly illustrates the scale of the problem under consideration. 

Given that the `disappearance' of such a large fraction of the available oxygen needs to be accounted for, there are perhaps only a very limited number of possible or viable solutions to this problem, namely:
\begin{enumerate}
\item The derived O abundances are over-estimated: an O abundance of $\simeq 400$\,ppm rather than the currently accepted values of 490 and 575\,ppm could practically resolve the problem. However, this seems highly improbable given the extensive and very careful work that has been done to determine the cosmic O abundance \citep{2008ApJ...688L.103P,2009ARA&A..47..481A,2009ApJ...700.1299J}. 
\item There exists an unobservable, chemically-peculiar, IR-inactive, O-rich material, which must be associated with other abundant elements.  This would appear to be a particularly unphysical solution. 
\item Trapping the missing O atoms in carbonyl species (C$=$O) is possible but it would require depleting an equal quantity of carbon into this carbonyl phase, which ought to have detectable IR absorption bands (see Sect. \ref{sect_spectra}), and is therefore seemingly untenable. 
\item Sequestering the missing O atoms into an organic carbonate ($^{\rm -O}_{\rm -O}$$>$C$=$O), using the most likely reacting/accreting partner carbon, would maximise the oxygen/carbon ratio and could therefore be a viable solution, provided that such a carbonate does not have strong, signature IR absorption bands that conflict with observations (see Sect. \ref{sect_spectra}).
\end{enumerate}

A self-consistent solution to this problem, in terms of O atom and CO molecule sequestration from the interstellar gas onto reactive nanoparticle surfaces was recently proposed \citep{2016RSOS....360224J}. This involves the formation of grain surface epoxide functional groups before the onset of ice mantle formation in translucent clouds, {\it i.e.}, in the transition between the diffuse and dense ISM. Such surface epoxides could be implicated in the formation of grain-surface oxygen-rich functional groups, {\it e.g.},  ketones, aldehydes, carboxylic acids, carboxylates and organic carbonates \citep{2016RSOS....360224J}, which perhaps ought to be detectable in translucent clouds, {\it i.e.}, before or at about the onset of ice mantle formation \citep{2016RSOS....360224J}.

\subsection{Abundance considerations}

Clearly, the only element that is abundant enough and that has the right chemical affinity for bonding to oxygen is carbon.\footnote{We can exclude hydrogen in this context because it is most likely to form water ice with oxygen and interstellar water ice abundances are well determined and deemed insufficient to explain the observed O depletion \citep{2010ApJ...710.1009W}.} We therefore now explore the viability and manifold consequences of depleting O atoms onto a-C(:H) dust as an organic carbonate. In Table~\ref{tab_O_budget} we indicate the elemental compositional constraints of this supposition. 

Respecting the cosmic abundances of the elements, and considering the lower limit O abundance of 490\,ppm, the stoichiometry of the O-trapping phase, normalised to carbon, is CO$_{x \simeq 0.1-0.5}$N$_{y \simeq 0.1}$H$_{z}$, where we now include nitrogen as an accreting and chemically-integrating heteroatom because it is observed in abundance in cometary organics and IDPs \citep[{\it e.g.},][]{2006Sci...314.1439N,2009LPI....40.1130D,2009LPI....40.2260D,2015Sci...349b0689G,2018PNAS..115.6608I}.  Here the lower limits of 0.1 for $x$ and $y$ come from the typical O and N abundances observed in cometary and IDP organic carbons \citep[{\it e.g.},][]{2018PNAS..115.6608I} and the upper limit of $x \simeq 0.5$ from the requirement to deplete all of the gas phase oxygen, not in silicates, CO and water ice, in dense regions \citep[{\it e.g.},][]{2010ApJ...710.1009W}, {\it i.e.}, from Table~\ref{tab_O_budget} we have O/C $= (30+140) / (293+47) = 0.5$. The above-derived stoichiometry implies a homogeneous material with C:O $\simeq$ 1:0.5 throughout but if we instead assume a segregation into a meteoritic-like or IDP-like organic carbon phase with O and N atomic fractions of the order of 0.1 \citep[{\it e.g.},][]{2018PNAS..115.6608I} and also an organic carbonate-rich phase then we have:  
\[
0.86 \, ( \, {\rm CO}_{0.1}{\rm N}_{0.1}{\rm H}_{n} \, ) +  0.14 \, ( \, {\rm CO}_{3}{\rm H}_{m} \, ), 
\]
which yields the stoichiometry CO$_{0.51}$N$_{0.09}$H$_{z}$, where $z \simeq ( 0.86 n + 0.14 m )$. This allows for a relatively small fraction of carbon ($\simeq 12$\,\% $\equiv 47$\,ppm) to be associated with an O-rich organic material, for a cosmic O abundance of 490\,ppm. 

If instead we adopt the higher O abundance of 575\,ppm the overall stoichiometry becomes CO$_{x \simeq 0.1-0.75}$N$_{y \simeq 0.1}$H$_{z}$, where from Table~\ref{tab_O_budget} the upper limit on $x$ is given by $ (27+228) / (264+76) = 0.75$. Upon decomposition (as above) this yields: 
\[
0.78 \, ( \, {\rm CO}_{0.1}{\rm N}_{0.1}{\rm H}_{n} \, ) +  0.22 \, ( \, {\rm CO}_{3}{\rm H}_{m} \, ), 
\]
which requires $\approx 60$\,\% more carbon ($\simeq 19$\,\% $\equiv 76$\,ppm) to be incorporated into an organic carbonate phase. While the latter high abundance of carbon trapped into a carbonate material is indeed possible it is obviously more problematic because it requires depleting so much more carbon into this dust phase. 

\begin{table}
\caption{The ISM oxygen budget redistribution in dense clouds, out of a total abundance of (490:575) (\cite{2009ARA&A..47..481A}:\cite{2008ApJ...688L.103P}), where these values are used to determine the requisite (lower\,:\,upper) O and (upper\,:\,lower) C depletions into  a-C:H and carbonate dust phases. Column 3 shows the equivalent carbon abundances out of a budget of 390 ppm \citep{2009ApJ...700.1299J,2012ApJ...760...36P}. Column 4 indicates a possible nitrogen depletion into dust. The gas phase CO and ice phase O depletions are taken from \cite{2010ApJ...710.1009W}. a-C(:H)[:O:N] represents an a-C:H solid phase with $\simeq 10$ atomic\,\% O and N heteroatoms present in carbonyl (C$=$O), aldehyde ($-$C$\leqslant^{\rm H}_{\rm O}$), ketone ($>$C$=$O), amide ($-$C$\leqslant^{\rm NH_2}_{\rm O}$), imine ($>$C$=$N) and nitrile ($-$C$\equiv$N) functional groups, as per the GEMS-containing IDPs analysed by \cite{2018PNAS..115.6608I}.}             
\label{tab_O_budget}      
\centering                          
\begin{tabular}{l c c c}        
\hline\hline \\[-0.25 cm]                 
                &  O         & C    & N   \\    
   Phase   &  [ppm]   & [ppm]  & [ppm]   \\[0.1cm]    
\hline \\[-0.25 cm]                        
    a-Sil$_{\rm (s)}$                       &  110  &  --- &  ---  \\      
    CO$_{\rm (g)}$                         &    50  &   50 &  --- \\      
    H$_2$O$_{\rm (s)}$-ice           &  160  &  ---  &  ---  \\      
    a-C, a-C(:H)[:O:N]$_{\rm (s)}$   &    (30:27)  &  (293:264) &  30 \\[0.1cm]      
    organic carbonate$_{\rm (s)}$   &    $\geq$ (140:228)  & $\gtrsim$ (47:76)  & ---  \\
      \ \ \ \ ( O:C = 3:1 )                    &       &    &  \\[0.1cm]      
    \hline \\[-0.25 cm]                                   
       Total          &    $\geq$ (490:575)   &  $\gtrsim 390$ &  30  \\[0.1cm]      
    \hline                                   
\end{tabular}
\end{table}

\subsection{Spectroscopic considerations}
\label{sect_spectra}

The proposed oxygen-trapping hypothesis requires incorporating $\simeq 35-44$\,\% of the cosmically-available oxygen, combined with $\gtrsim 12-19$\,\% of the carbon, in a-C(:H)[:O:N] mantles and organic carbonate in dense clouds \citep[assuming $\simeq 50$\,ppm of carbon in CO molecules,][]{2010ApJ...710.1009W}.  We now consider the detectability of such a carbonate-rich phase, which must clearly exhibit detectable IR absorption features due to the presence of polarised carbon-oxygen bonds.  

Manufactured polycarbonate polymers consist of cross-linked, ordered linear chains that result in very durable materials. They clearly do not absorb at visible wavelengths (unless a colourant has been added) but do absorb in the UV. However, they do not degrade sufficiently fast  in response to solar UV to limit their accumulation ({\it viz.}, pollution) and their contribution to the floating plastic continents of the Earth's oceans. Industrial organic polycarbonates are synthesised  by the reaction of carbon dioxide with epoxides, a route that readily forms cyclic organic carbonates. In order to minimise cyclic carbonate formation, ensure quality and a high yield ($\geqslant$ 50\%) an organic ligand, bimetallic catalyst is required in the industrial process \citep[{\it e.g.},][]{organic_polycarbonate_formation}. It therefore appears that cyclic organic carbonates form readily but must be avoided  in the manufacture of polycarbonate plastics, where they would disrupt and weaken the polymer integrity. 

The five-membered (or more) cyclic organic carbonate structure (indicated by the prefix c--) can be represented as:
\[
{\rm c-} \left( [^{\rm C-O}_{\rm C-O}\hspace{-0.0cm}>\hspace{-0.1cm}{\rm C}\hspace{-0.1cm}=\hspace{-0.1cm}{\rm O}  \right), 
\]
where the [$^{\rm C}_{\rm C}$ part of the structure represents a contiguous part of an a-C(:H) grain surface to which a carbonate functional group, $^{\rm -O}_{\rm -O}$$>$C$=$O, is bonded and that is therefore not counted in the carbon budget of the organic carbonate because it is an integral part of the grain. 

Well-ordered industrial polycarbonates show distinct IR absorption bands at $\simeq 5.6\,\mu$m (carbonyl C=O stretching), $\sim 8.1-8.6\,\mu$m  (O--C--O asymmetric stretching) and $\simeq 9.5\,\mu$m (O--C--O symmetric stretching).  The characteristic and identifying IR absorption bands of cyclic organic carbonates occur at $\sim 5.5-5.8\,\mu$m and $\sim 7.8-8.2\,\mu$m IR wavelengths\footnote{For comparison, the mineral carbonates found in carbonaceous chondritic meteorites have characteristic IR bands in the $6.67-7.69\,\mu$m wavelength region.} \cite[{\it e.g.},][and references therein]{2016RSOS....360224J} and while they do show other IR-active modes these are not unique to them. Indeed both cyclic organic carbonate bands could be merged with the observed aromatic carbon emission bands, which peak in the $6.2-8.6\,\mu$m wavelength region \citep[{\it e.g.}, at $\simeq 6.2, 7.5-7.8$ and $8.6\,\mu$m,][]{2001A&A...372..981V}, and/or the wing of the $\simeq 9.7\,\mu$m amorphous silicate Si--O stretching band generally observed in absorption.  As close inspection shows, the cyclic organic carbonate bands could be confused with this  blue wing and, if they exist in interstellar spectra, also with the sub-bands recorded in some of the most recent amorphous, Mg-rich, silicate laboratory data that show a distinct band at $\sim 6\,\mu$m and a shoulder on the Si--O stretching band at $\sim 8\,\mu$m  \citep{2017A&A...606A..50D,2017A&A...600A.123D}. It should be noted that the $5.5-5.8\,\mu$m band generally lies in a less-confused but seemingly little observed part of the IR spectrum. In a number of luminous infrared galaxies, with abundant a-C(:H) dust, a band near $5.8\,\mu$m is indeed observed, which is suggestive of a carbonyl bond absorption in ketones and/or aldehydes \citep{2004A&A...423..549D,2007A&A...463..635D,2007A&A...476.1235D}. 

\cite{2000ApJ...537..749C} observed an absorption band at $\sim 6\,\mu$m along all the lines of sight towards the Galactic Centre that they studied. They concluded that this feature cannot be due to water ice alone and that, even with a contribution from a WR-type $6.2\,\mu$m aromatic carbon band, a small amount of the carboxylic acid HC$\leqslant^{\rm OH}_{\rm O}$ (formic acid) is required in the mix to fit the observed ice band along the line of sight towards Sgr~A\textasteriskcentered. 

In a study of the ice absorption features in the $5-8\,\mu$m region along the lines of sight towards ten embedded protostars \cite{2001A&A...376..254K} find that spectra are dominated by $6.0\,\mu$m and $6.8\,\mu$m  absorption bands, which vary significantly among the studied lines of sight. They find that the $6.0\,\mu$m absorption band can principally be assigned to amorphous water ice but towards a number of sources an additional absorption  at $5.83\,\mu$m, due to formic acid, HC$\leqslant^{\rm OH}_{\rm O}$, and at $5.81\,\mu$m due to formaldehyde, HC$\leqslant^{\rm H}_{\rm O}$, is required, in addition to a $6.2\,\mu$m aromatic C--C stretching component.  

\cite{2015ARA&A..53..541B}, in their ``Icy Universe'' review, discuss the decomposition of the observed interstellar ice bands in the $5-8\,\mu$m region, which in a five component fit requires a broad band peaking at $5.9\,\mu$m (denoted C5) underlying the entire $5-8\,\mu$m range. The C5 feature appears to be decoupled from the shorter wavelength bands and is probably due to some as-yet undetermined ice component. However, the width of the C5 feature and confusion with the H$_2$O ice bending mode at $6\,\mu$m makes it hard to accurately quantify the variations in this band \citep{2015ARA&A..53..541B}. 

Seemingly, the interpretation of the `ice' absorption features in the $5-8\,\mu$m wavelength region is not yet clear-cut and so there is probably some room for complementary explanations. For instance, it is possible that the broad $5.9\,\mu$m (C5) band \citep{2015ARA&A..53..541B} and the $5.8-6.0\,\mu$m absorption feature attributed to carbonyl-containing aldehyde and carboxylic acid \citep{2000ApJ...537..749C,2001A&A...376..254K} could be organic carbonate signatures, rather than a blend of  carbonyl and carboxyl functional groups. We note that the IR vibrational modes of organic carbonate functional groups in an amorphous hydrocarbon, a-C(:H), solid are likely to be perturbed and therefore rather broad. 

We now examine the question of the detectability of carbonyl bands, be they components of aldehydes, carboxylic acids, carboxylates or carbonates, in a little more detail. 

\cite{2004A&A...423..549D} observed the signatures of the organic matter in the diffuse ISM of the Seyfert galaxy NGC\,1068 and compared them with the same bands along lines of sight through the Milky Way towards the Galactic Centre region. From their coupled observational and experimental analysis of the NGC\,1068 data \cite{2004A&A...423..549D} conclude that the elemental composition of the diffuse ISM carbonaceous matter, a-C(:H)[:O], exhibiting a C=O carbonyl band at $5.87\,\mu$m, has a carbon to oxygen ratio C/O $\simeq 9$. This is the same value found by \cite{2018PNAS..115.6608I} for the average composition of the organic carbon matrix in one of their IDPs ({\it i.e.}, C$_{83}$N$_{8}$O$_{9}$). Further, and using their NGC\,1068 data as a model, \cite{2004A&A...423..549D} determined that, for oxygen in carbonyl bonds and carbon in aliphatic form, the O/C ratio that could escape detection, if it were to be present in the carbonaceous matter towards the Galactic Centre, is O$_{\rm (carbonyl)}$/C$_{\rm (aliphatic)} \lesssim 40$. For the organic carbonate oxygen trap proposed here (see Table \ref{tab_O_budget}) the predicted O$_{\rm (carbonyl)}$/C$_{\rm (aliphatic)}$ ratio is of the order of 2 to 3, adopting the THEMIS model aliphatic carbon fraction of 0.25 \citep{2017A&A...602A..46J}. Thus, and by analogy with carbonyl band observations, it appears that it would indeed be extremely difficult to observe the IR absorption signatures of a large fraction of oxygen if it were to be trapped in the form of an organic carbonate. This holds true even if the organic carbonate IR signature bands were to be an order of magnitude more intense than the carbonyl bonds in the ketones that were investigated by \cite{2004A&A...423..549D} in their comparative study. 

The utility of x-ray absorption spectroscopy in determining the composition of refractory interstellar dust has been quite remarkable \citep[{\it e.g.},][]{2005A&A...444..187C,2012A&A...539A..32C,2011ApJ...738...78X}. Further, laboratory studies of primitive solar system organic materials using x-ray absorption near-edge structure (XANES) spectroscopy reveal their rich and divers chemical functionality \citep[{\it e.g.},][]{2011M&PS...46.1376D,2011LPI....42.2603D}.\footnote{Table 2 of this work is a useful compilation of the characteristic carbon and nitrogen XANES spectra transition energies in organics. This compilation also indicates that aragonite, a crystalline calcium carbonate mineral, exhibits an edge feature at 0.2903\,keV.}  Thus, and in principle, x-ray absorption spectroscopy could also be applied to explore the composition of organic matter in space, including organic carbonates, through their near edge x-ray absorption fine structure features ({\it e.g.}, via NEXAFS, XANES, \ldots). However, it is currently not possible to explore carbon K-edge features ($E_{h\nu} > 0.28$\,keV)  with the high-resolution spectral energy coverage provided by x-ray telescopes ($E_{h\nu} \gtrsim 0.3$\,keV) because the detectors are often carbon-based. The only immediate hope is then to search for oxygen near edge x-ray absorption features (oxygen K-edge, 0.538\,keV) characteristic of organic matter, which would have to be differentiated and discriminated from the likely dominant features due to oxygen in refractory minerals (silicate, oxides, \ldots). Further, given that the oxygen edge in Mg-rich silicates is similar to that of water ice \citep{2012A&A...539A..32C}, it is clear that extracting interstellar organic matter absorption edge features due to oxygen-containing functional groups is likely to prove difficult. In conclusion, x-ray observations are unfortunately probably not yet mature enough to be able to discern the nature of any type of interstellar organic matter, let alone detect interstellar organic carbonate. However, if and when they do become available x-ray absorption observations of the carbon and oxygen K-edges ($E_{h\nu} \gtrsim 0.2$\,keV) will provide both a qualitative (test of the hypothesis) and a quantitative measure of any organic carbonate in the ISM.

\section{Discussion}
\label{sect_discuss}

Given the uncertainty ($\simeq 16$\,\%) in the cosmic abundance of oxygen, {\it i.e.}, $\simeq 490-575$\,ppm, and on the exact composition,  likely stability and reactivity of an O-rich, CO$_x$H$_n$, solid phase, we now consider the evidence and some of the wider implications relating to organic carbonates in the ISM.

\subsection{(In)Direct evidence}
\label{sect_discuss_1}

The observational evidence clearly indicates that oxygen undergoes significant depletion \citep{2009ApJ...700.1299J} before the onset of ice mantle formation and that this must occur into some form that is not ice-like \citep{2010ApJ...710.1009W}. 

Primitive solar system bodies, such as comets, meteorites and IDPs, ought to provide relevant evidence in the sense that they represent dense cloud matter that was subsequently exposed to star and planet formation. Hence, these primitive solar system bodies must carry a vestige of the chemical make-up of the dense, star-forming cloud that gave birth to the Solar System. However, to date no direct analyses of the volatile icy materials, which could contain vestiges of the dense cloud matter that predated the formation of the Solar System, has been possible.  

Turning our attention first to comets. The results of the COmetary Sampling And Composition (COSAC) instrument, onboard the Rosetta/Phil\ae\ mission to comet 67P/Churyumov-Gerasimenko, have taught us an enormous amount about the chemical composition of the volatile components of this comet. However, the detected species have most probably already undergone some processing before detection. In particular, carbonyl ($>$C=O) species are notably abundant among the detected organic molecules which include:  aldehydes, R--C$\leqslant ^{\rm H}_{\rm O}$ (R = CH$_3$, --CH$_2$OH and --CH$_2$CH$_3$), amides, R--C$\leqslant ^{\rm NH_2}_{\rm O}$ (R = H and CH$_3$) and the ketone (CH$_3$)$_2$C=O \citep{2015Sci...349b0689G}. Despite a very rich chemistry it was rather surprising that ammonia, NH$_3$, methanol, CH$_3$OH, carbon dioxide, CO$_2$ and carboxylic acids, R--C$\leqslant ^{\rm OH}_{\rm O}$ were not detected in measurable quantities by COSAC but were, apart from the latter, observed in the coma by the ROSINA mass spectrometer onboard Rosetta \citep{2016E&PSL.441...91M}. However, these coma species could be the photolytic break-down, daughter products of more complex molecules that were released into the coma. Perhaps a bigger surprise from the COSAC results was that formamide, H--C$\leqslant ^{\rm NH_2}_{\rm O}$, was the most abundant molecule after water, at a concentration $\sim 2$\,\% relative to water. It is worthy of note that half of the detected species contain C=O bonds, in $>$C=O and =C=O functional groups, and half (but not the same half) contain nitrogen atoms in amine (--NH$_2$), imine (--N=) and nitrile (--C$\equiv$N) functional groups. 
 
We now consider meteorites. The least-thermally processed organic matter in some carbonaceous chondrites contains fewer aromatic C=C bonds, more nitrogen and a higher ketone ($>$C=O) and carbonyl (C=O) fraction than the more aromatic organic phase \citep{2015LPICo1856.5128D}. This anti-correlation between aromatic content and oxygen-containing functional groups is consistent with parent body heating which renders carbonaceous matter more aromatic and drives out O-rich functional groups on route to forming a poorly graphitised type of carbon \citep{2015LPICo1856.5128D}. In a similar vein, a study and comparison of the carbonaceous components of cometary, IDP, meteoritic and diffuse ISM matter, \cite{Matrajt:2013ky} found that comet Wild~2 and ultracarbonaceous IDP particles are richer in aliphatic CH$_2$ and carbonyl (C=O) groups than that in the insoluble organic matter (IOM) in meteorites and in diffuse ISM dust. 

Finally we draw on some recent analyses of two cometary IDPs by \cite{2018PNAS..115.6608I}, which are equal mixtures by volume ($\equiv 20/80$\,\% by mass) of organic carbons and a-Sil GEMS (Glass with Embedded Metal and Sulphides).  The organic matter has an average composition of $\approx$ C$_{10}$NO and exists in two forms: an aliphatic, low-density matrix and a higher density more aromatic material present as mantles on GEMS grains, their sub-grains and as blobs within the lower-density, matrix carbon. Also present are higher density organic nano-globules. The elemental and isotopic compositions of the analysed organic matter in the IDPs points to its formation in cold environments \citep{2018PNAS..115.6608I}, such as dense clouds.

\subsection{Thoughts and speculations}
\label{sect_discuss_2}

For $\simeq 12-19$\,\% of all carbon to be incorporated into an organic carbonate, as proposed in the previous section, would require a solid material with a C:O ratio of 1:3. If instead the material is less O-rich, {\it e.g.}, for a stoichiometric ratio of 1:2 as in a carboxylate, $-$C$\leqslant^{\rm O-}_{\rm O}$ (or CO$_2$), or 1:1 as in a carbonyl, C$=$O, this would require tying up $\simeq 18-28$\,\% and $\simeq 36-57$\,\% of the carbon, respectively, depending on the cosmic abundance of oxygen. Clearly if, in the transition from the diffuse to the dense ISM, oxygen depletes into a material with a stoichiometry other than the most O-rich possible, {\it i.e.}, an organic carbonate (CO$_3$), this would tie up an uncomfortably large fraction, {\it i.e.}, $\approx 20-60$\,\%, of the available carbon in a single material, which ought then to be observable but is probably not as argued above in Section \ref{sect_spectra}. 

If the formation of an organic carbonate phase in dust is the only viable solution to the oxygen depletion problem, the questions then become: how and where could such an organic carbonate phase form and incorporate into dust? And why does this phase not seem to persist? Organic carbonate formation must obviously occur through the accretion of oxygen-carrying species ({\it e.g.}, O, OH, CO, \ldots) and be driven by the increased density and attenuation of the medium in which it occurs, {\it i.e.}, increased with respect to the diffuse ISM and therefore most likely to occur in translucent clouds.  

It has been noted that reactive epoxides, $>$C$_{-}^{\rm O}$C$<$,\footnote{And also their nitrogen and sulphur analogues, aziridine, $>$C$_{-}^{\rm N}$C$<$, and episulphide, $>$C$_{-}^{\rm S}$C$<$.} are as-yet under-explored chemical reagents in interstellar chemistry  \citep{2016RSOS....360221J,2016RSOS....360224J}. They readily react with CO$_{2 \rm (g)}$ to form organic carbonates\footnote{Note that the reaction of CO$_2$ with epoxide is one of the industrial routes to polycarbonate plastics.} and must react with CO$_{\rm (g)}$ in a similar way to form carbonyl bonds in species such as ketones, carboxylates and carboxylic acids, in addition to organic carbonates \citep{2016RSOS....360224J}. Thus, epoxide chemistry would provide a chemically-viable route to organic carbonate formation and indeed to a very rich chemistry in general.

\subsection{Organic carbonate formation and photolysis}
\label{sect_discuss_3}

In the ISM the formation of organic carbonates and carbonyl-rich species is most likely to occur in regions somewhat denser and more UV-shielded ($A_{\rm V} \gtrsim 1$) than those where surface epoxides could actively be driving OH formation \citep{2016RSOS....360221J}, and where CO, C and O are still abundant in the gas. Further, as the observational evidence indicates, this must occur before ice mantle accretion \citep{2010ApJ...710.1009W}, which occurs for through-cloud $A_{\rm V} \geqslant 3.2\pm0.1$ \citep[{\it e.g.},][]{2015ARA&A..53..541B} in the ISM but this threshold is somewhat variable \citep{2001ApJ...547..872W}. Carbonate formation would therefore likely occur in translucent clouds (Table~\ref{table_ISM}) with into-cloud optical depths $A_{\rm V} \simeq 1-1.6$ and  should therefore occur at about the same time as a-C:H mantles accrete \citep{Faraday_Disc_paper_2014,2015A&A...579A..15K,2016A&A...588A..43J,2016A&A...588A..44Y,2016RSOS....360224J} and a-C nanoparticles coagulate onto large grain surfaces in the transition from the diffuse to dense ISM \citep{Faraday_Disc_paper_2014,2015A&A...579A..15K}. This will, after over-lying ice mantle accretion, lead to the kinds of complex grain composition and structures indicated on the left hand side of Fig.~\ref{fig_essentials}, where coagulation between big grains could also occur but was not included in this figure for clarity. 

Having considered the grain surface formation of O-rich organic carbonates, carboxylates and carbonyls, we now turn our attention to what happens to them when grain mantles are UV photolysed during star formation and in PDRs. As noted  above the anti-correlation between aromatic content and oxygen-containing functional groups, in Solar System organics, implies this O-rich dust mantle component is depleted during photolysis \citep[{\it e.g.},][]{2015LPICo1856.5128D}, which is coherent with the results of \cite{Matrajt:2013ky}. The UV photon-driven break-down of organic carbonates is likely to yield daughter products that include, at least, CO, CO$_2$ and aldehydes ($-$C$\leqslant^{\rm H}_{\rm O}$) \citep{back_parsons_1983}, and could possibly also yield ketones ($>$C$=$O), carboxylates ($-$C$\leqslant^{\rm O-}_{\rm O}$) and carboxylic acids ($-$C$\leqslant^{\rm OH}_{\rm O}$). The UV photolysis of the a-C(:H)[:O:N] solid phase, with $\simeq 10$ atomic\,\% O and N heteroatoms, {\it i.e.}, the [:O:N] component, as measured in meteoritic, IDP and cometary organics, will yield some of the above products but could, as we speculate here, also lead to amides ($-$C$\leqslant^{\rm NH_2}_{\rm O}$), imines ($>$C$=$N) and nitriles ($-$C$\equiv$N). The effects of UV photolysis on the dust in transition regions is schematically illustrated in Fig.~\ref{fig_essentials}, where mantle mass loss through de-volatilisation and some likely daughter photo-desorption and photo-dissociation products are indicated.

\begin{table}
\caption{The possible functional groups for O and N heteroatoms incorporated into the aliphatic, olefinic and aromatic components of aliphatic-rich, a-C:H, and aromatic-rich, a-C, phases in (hydrogenated) amorphous carbon, a-C(:H), materials.} 
\label{table}      
\centering                          
\begin{tabular}{c c l}        
\hline\hline \\[-0.25 cm]                 
          & functional &  \\    
 element  & group   & \\[0.1 cm]    
\hline \\[-0.25 cm]                        
 N & $\geqslant$C$-$N$<$ & aromatic   \\[0.1 cm]      
 N & $>$C=N--  & imine    \\[0.1 cm]      
 N & $>$C=NH & imine       \\[0.1 cm]      
 N & --NH$_2$ & amine  \\[0.1 cm]      
 \hline  \\[-0.25 cm]                                  
 O & {\Large $\hexagon$}\hspace{-0.21cm}{$^{\rm O}$} & aromatic ketone         \\[0.1 cm]      
 O & $>$C=O &  ketone       \\[0.1 cm]      
 O & {\tiny $\diagup$}$^{\rm O}${\tiny $\diagdown$} & ether               \\[0.1 cm]      
 O & {\Large $\vartriangle$}\hspace{-0.25cm}$^{\rm O}$ & epoxide           \\[0.1 cm]      
 O & --OH  & alcohol            \\[0.1 cm]      
 \hline   \\[-0.25 cm]                                 
\end{tabular}
\label{table_func_groups}
\end{table}

\subsection{Organic carbonate and mantle evolution}
\label{sect_discuss_4}

The evolution of O, and also of N, heteroatoms within amorphous (hydro-)carbon materials is interesting and key to this discussion. These two elements tend to behave differently within these materials because of their particular bonding configurations and the differing chemical affinities/reactivities and photo-sensitivities. While both elements can incorporate into aliphatic and olefinic structures (see Table \ref{table_func_groups}), and are the most common heteroatoms in aromatic systems, nitrogen is more prevalent in aromatic systems. This is because it can substitute for a carbon atom and preserve the ring aromaticity, while oxygen atoms can only do so as O$^+$, which is iso-electronic with an N atom. Oxygen atoms are therefore less likely than nitrogen atoms to be incorporated into aromatic ring systems during the UV photolysis, and concomitant aromatisation, of aliphatic-olefinic a-C(:H)[:O:N] mantles. Further, grain surface oxygen heteroatoms may be more likely to react with incident H atoms in the ISM, leading to gas phase species, than are nitrogen atoms. This general idea would appear to be consistent with the analysis of the organics in meteorites, comets and IDPs \citep[{\it e.g.},][]{2015LPICo1856.5128D,2018PNAS..115.6608I,Matrajt:2013ky}, where O and N atoms are practically equally abundant despite the much higher cosmic abundance of oxygen. This difference can be attributed to the fact that nitrogen atoms have a stronger affinity for and stability within aromatic phases, than oxygen, and are therefore more stable against thermal processing and UV photolysis. In aromatics  nitrogen atoms can readily substitute for carbon and in these configurations are hard to detect because of the close frequencies of the IR active C=C and C=N bond modes \citep{2004PhilTransRSocLondA..362.2477F}. Hence, the more processed aromatic-rich organics are enhanced in nitrogen (present as $>$N-- and =N--), with respect to oxygen, while the aliphatic organics should be more carbonyl-rich (with $>$C=O, present as aldehydes, ketones, carboxylates, carbonates, \ldots), which again appears to be coherent with meteorite, comets and IDP studies \citep[{\it e.g.},][]{2015LPICo1856.5128D,2018PNAS..115.6608I,Matrajt:2013ky}. 

As pointed out above organic carbonates would form before icy mantles accumulate and, at the other end of the evolutionary cycle, be processed after most or all of the icy mantles are removed from the grains surfaces. Hence, and because of this decoupling of surface-bonded carbonates from the more volatile, molecular ice mantles (H$_2$O, CO$_2$, CO, CH$_3$OH, \ldots), we do not consider the chemical effects of these icy mantles, which have been discussed elsewhere within the context of the THEMIS dust model \citep{2015A&A...579A..15K,2016RSOS....360224J,2018arXiv180410628J}. 

As is well known, carbonates of organic, inorganic or mineral composition thermally decompose to an oxide and release CO$_2$. Experimentally, it has been shown that the thermal and UV photolytic induced decomposition of the organic carbonate molecule ethylene oxide (oxirane, see Fig.~\ref{fig_chemistry}) yields CO$_2$, ethylene oxide and H$_2$, CO, CH$_4$ and CH$_3$C$\leqslant^{\rm H}_{\rm O}$ as minor products  \citep{back_parsons_1983}. Thus, it is entirely feasible that the UV photolysis of an organic carbonate in the ISM would release CO$_2$ (to the gas), form residual surface epoxide and carbonyl functional groups and also release other daughter products, such as CO, CH$_n$ and CH$_n$C$\leqslant^{\rm H}_{\rm O}$, in the process (see Fig.~\ref{fig_essentials}, right hand side, and Fig.~\ref{fig_chemistry}), {\it i.e.}, 
\[
[^{\rm C-O}_{\rm C-O}>\hspace{-0.1cm}{\rm C}\hspace{-0.1cm}=\hspace{-0.1cm}{\rm O}
\hspace{0.5cm} \rightarrow \hspace{0.5cm} 
[^{\rm C-}_{\rm C=O} \hspace{0.5cm} {\rm or} \hspace{0.5cm} [^{\rm C}_{\rm C}>\hspace{-0.1cm}O  \hspace{0.2cm} + \hspace{0.2cm} CO_{\rm 2 (g)}^\uparrow
\]
\vspace*{-0.5 cm}
\[
{\rm carbonate \hspace{1.4cm} carbonyl \hspace{0.9cm} epoxide}
\]
This probably occurs at about the same time as the aliphatic-rich a-C:H mantles are photolytically de-volatilised,  {\it i.e.}, likely where $A_{\rm V} \lesssim 1$, $n_{\rm H} \lesssim 10^3$\,cm$^{-3}$ and $G_0 \gg 1$ \cite[{\it e.g.},][]{2014A&A...569A.119A,2015A&A...584A.123A}. 

That carbonates in the ISM will decompose to release CO$_2$ may have some bearing on the problematic formation of CO$_2$ in interstellar ices \citep{1997ApJ...490..729W,1998ApJ...498L.159W,2007ApJ...655..332W,2009ApJ...695...94W}, as discussed by \cite{2016RSOS....360224J}.   

\begin{figure*}
\centering
\includegraphics[width=18cm]{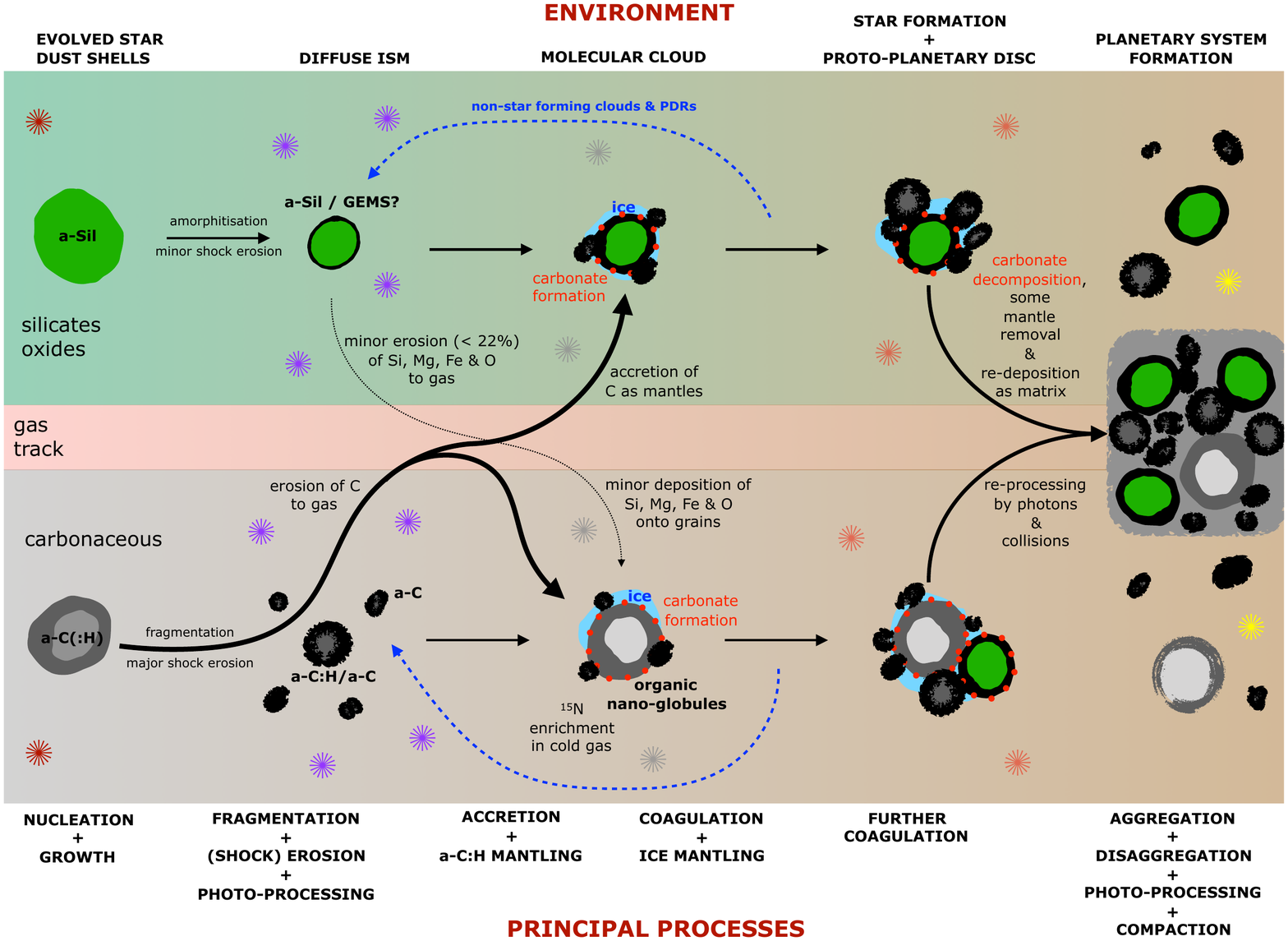}
\caption{The evolutionary flow of dust through the ISM, from its formation around evolved stars (left) via diffuse and dense clouds (middle) towards star formation and planetary systems (right).}
\label{fig_dust_flow}
\end{figure*}

Carbonaceous organic matter processing or maturation is likely to follow something along the lines of the (condensed) sequence described below, which is based upon the discussions of \cite{2016RSOS....360224J}, is illustrated in Fig.~\ref{fig_essentials} and in more detail within a global dust evolution framework in Fig.~\ref{fig_dust_flow}, {\it i.e.},  
\begin{enumerate}
\item Accretion of O, CO and N mono- or multi-layers, before ice mantle formation, leading to the formation of carbonyl-rich organics, including carbonates, through reactions at epoxide-functionalised grain surfaces. 
\item Accretion of a-C:H[:O:N]  mantles rich in oxygen and nitrogen, with [O] $\gg$ [N], in the form of carbonyl and N-containing functional group species, including: aldehydes, ketones, carboxylates, organic carbonate, amides, amines and imines. 
\item UV photolysis drives the decomposition of carbonates and release of CO$_2$, and the partial dehydrogenation of aliphatic-rich, carbonyl-rich a-C:H[:O:N] to a material with [O] $\gtrsim$ [N]. 
\item Extended UV photolysis leads to the decomposition of carbonyl groups, the further loss of other oxygen-containing groups and to the aromatic enrichment of a-C(:H)[:O:N] mantles to materials with [O] $\sim$ [N].  
\end{enumerate}
During this process, and while O and N atoms are lost from the material, N atoms are sequestered into aromatic structures which are more resistant to photolysis than carbonyl-containing functional groups, which leads to an equilibriation of the O and N atom concentrations compatible with IDP studies \citep[{\it e.g.},][]{2018PNAS..115.6608I}. 

Fig.~\ref{fig_dust_flow} gives a schematic overview of the evolutionary flow of dust from its formation around evolved stars, its passage through the diffuse ISM, translucent clouds, on into dense clouds and finally towards star formation and planetary system formation. Of particular note is the  relatively small increase in the observed Si depletion, from 27 to 33\,ppm ($\equiv 78$ to 96\,\%), with increasing fraction of molecular hydrogen, $f({\rm H_2})$, from 0.01 to 1 \citep{2016AJ....151..143H}, which is indicated in Fig.~\ref{fig_dust_flow}.\footnote{{\it N.B.} The THEMIS model uses 80\,\% of the available Si in its a-Sil dust component, which compares well with the observed value of 78\,\% \citep{2016AJ....151..143H}.} This observational result implies that silicon is a rather resistant dust element that is predominantly trapped within dust even in rather extreme environments \citep[{\it e.g.},][]{2002ApJ...579..304W,2016AJ....151..143H} and that it does not therefore need to be re-made in the ISM \citep{2011A&A...530A..44J}. 

Illustrated in Fig.~\ref{fig_dust_flow} is the transitory nature of the proposed organic carbonate phase, which is indicated by the appearance and disappearance of the red dots. Thus, this hypothesis predicts that the oxygen-trapping phase is rather transient, because of its limited stability to thermal processing and photolysis, and is therefore likely to be a difficult material to identify either spectroscopically or by direct in-situ analyses. However, it seems that many of the observed cometary organics are at least not inconsistent with the break-down daughter products of organic carbonates and a-C:H[:O][:N] solids and therefore, in principle, with the organic carbonate oxygen-trapping hypothesis proposed here.

\section{Organic carbonate elucidation strategies}
\label{sect_elucidation}

Given its rather ephemeral nature it is currently seemingly rather difficult to either confirm or refute the existence of significant quantities of organic carbonate in dust in interstellar media. Thus, it appears that new experimental and observational strategies will be needed in order to test this  hypothesis and to this end we suggest the following as possible routes to elucidation or invalidation: 
\begin{enumerate}

\item The fundamental chemistry of the interstellar organic carbonate hypothesis can be tested with a two-stage experiment: \\ 
A) O-atom reaction with a-C(:H) substrates to form epoxide functional groups (epoxylation), followed by \\ 
B) reaction of incident CO and CO$_2$ molecules with epoxylated a-C(:H) to form organic carbonyl and carbonate groups.\label{item_lab_test}

\item There is a clear need for more spectral data in the $4-8\,\mu$m window, both from laboratory experiments and astronomical observations. Indeed the above experimental procedure will need to be verified by spectroscopic measurements. 

\item The tell-tale spectroscopic signatures of organic carbonates ($\sim 5.5-5.8\,\mu$m and $\sim 7.8-8.2\,\mu$m) should be searched for in astronomical observations of transition regions, {\it i.e.}, translucent clouds and the outer edges of quiescent molecular clouds. The JWST-MIRI instrument would appear to be an ideal instrument with which to undertake this spectral exploration. However, \ldots

\item A careful characterisation of the short-wavelength wing of the amorphous silicate band at $\simeq 9.7\,\mu$m will be required in order to ensure that no silicate material, or other O-rich oxide mineral, spectral features  in the $\sim 5-9\,\mu$m region could be confused with the sought-for organic carbonate bands. 

\item A deeper search for the spectroscopic absorption and reflectance signatures of organic carbonates in cometary matter could also be undertaken once appropriate laboratory data become available (suggestion \ref{item_lab_test} above). 

\item The x-ray absorption features of organic carbonyl functional group-containing and cyclic organic carbonates should be laboratory-characterised ({\it e.g.}, via NEXAFS, XANES, \ldots \ for $E_{h\nu} > 0.2$\,keV) in order to determine whether it would be possible to detect them in carbon K-edge spectra and/or to distinguish them from interstellar oxygen K-edge features in silicate and oxide minerals and water ice.\footnote{{\it N.B.} Such experiments will need to avoid the breakdown of `fragile' organic matter during x-ray irradiation.} In the eventuality that they are detected these data would allow a quantification of the O and C abundances locked into organic carbonate. 

\item An observational search for organic (carbonate) matter in space, via x-ray K-edge absorption spectroscopy ($E_{h\nu} > 0.2$\,keV), should be made towards x-ray sources along lines of sight dominated by translucent or molecular clouds in regions where water ice mantles have not yet accumulated. Despite the fact that such ideal lines of sight may be a rather rare occurrence, these observations could prove absolutely conclusive. 

\end{enumerate}

\section{Conclusions}
\label{sect_concl}

In the transition from the diffuse through translucent and on into dense clouds of the order of one third to almost one half of the cosmic abundance of oxygen must accrete into an unknown, well-hidden, O-rich material.  

In order to most efficiently deplete oxygen without over-depleting carbon, its most likely elemental depleting partner, the best-case chemical composition would be that of an organic carbonate solid, $[^{\rm C-O}_{\rm C-O}>\hspace{-0.1cm}{\rm C}\hspace{-0.1cm}=\hspace{-0.1cm}{\rm O}$, which maximises the O/C atom ratio (3/1) and minimises the required carbon depletion into an O-rich phase.

Organic carbonates would need to form in the translucent ISM before the formation of icy mantles and at about the same time as aliphatic-rich a-C:H mantles accrete ({\it i.e.}, $A_{\rm V} \sim 1$). When these materials are illuminated and undergo UV photolysis during cloud collapse and star formation their decomposition products ought to resemble the observed cometary organic molecular make-up, as seen by the COSAC and ROSINA instruments on board the ROSETTA/Phil\ae\ mission to comet 67P/Churyumov-Gerasimenko. The thermal and UV photolytic decomposition of organic carbonates will liberate CO$_2$, which may have some bearing on the problematic formation of this molecule in ices in the denser ISM, something that has not yet been satisfactorily-resolved.  

During aliphatic-rich mantle accretion in the denser ISM nitrogen will also deplete into a-C:H, in addition to oxygen, but nitrogen will be preferentially incorporated into the more aromatic domains in organics during UV photolysis. Thus, mantles that may have a high O/N ratio in dense regions will evolve towards O/N $\sim 1$ as a result of processing, consistent with the nature of solar system organics. 

Much of the analysis and interpretation of so-called `ice' absorption features observed along interstellar lines of sight appears to be based upon the premise that the absorption bands are entirely due to a mix of molecular species. This is a fundamental and exclusive assumption 
that does not take into account the full range and the possible complexity of interstellar carbonaceous dust materials and their chemical reactivity with atoms, ions, molecules and radicals. This work shows that it might perhaps be more enlightening to drop this premise and to consider that the bands could, in part, also be due to macromolecular organic species. The cyclic organic carbonate proposed here is taken to be an integral, macromolecular part of an interstellar amorphous carbonaceous dust component, a-C(:H)[:O:N]$_{\rm(s)}$. A consideration of the possible  contributions of macromolecular materials  in the decomposition and interpretation of interstellar `ice' absorption features will necessarily complicate the experimental protocols and procedures necessary for the verification or negation of this hypothesis. Clearly these new avenues of possibility do need to be experimentally-explored.  

There are clearly great difficulties to surmount in the search for and elucidation (or invalidation) of any interstellar organic carbonate dust component. This is a frustration for the verification of the proposed interstellar (cyclic) organic carbonate hypothesis but these observational difficulties may also be a key factor in why, should they exist there, that organic carbonates have not already been detected in space. It is evident that an unequivocal identification of characteristic organic carbonate features, at IR and x-ray wavelengths, will prove a difficult task but one that is surely worth pursuing because it would yield significant insight into the fundamental role of gas-grain interactions in driving complex organic interstellar chemistry. 

\begin{acknowledgements}
The authors wish to thank their colleagues and collaborators for the many interesting discussions related to dust in interstellar media. 
We also wish to thank the anonymous referee for an encouraging report and some insightful remarks. 
\end{acknowledgements}

\bibliographystyle{aa} 
\bibliography{Ant_bibliography} 

\end{document}